\documentclass[fleqn,twoside]{article}
\usepackage{espcrc2}

\usepackage{graphicx}
\usepackage{epsfig}

\usepackage[figuresright]{rotating}

\newcommand{\AmS}{{\protect\the\textfont2
  A\kern-.1667em\lower.5ex\hbox{M}\kern-.125emS}}

\newcommand{\be}{\begin{equation}}
\newcommand{\ee}{\end{equation}}
\newcommand{\bi}{\begin{itemize}}
\newcommand{\ei}{\end{itemize}}
\newcommand{\ben}{\begin{enumerate}}
\newcommand{\een}{\end{enumerate}}
\newcommand{\bc}{\begin{center}}
\newcommand{\ec}{\end{center}}
\newcommand{\bea}{\begin{eqnarray}}
\newcommand{\eea}{\end{eqnarray}}

\def    \raw           {\rightarrow}

\def    \part          {\partial}

\def      \nue{\ensuremath{\nu_{e}}\ }

\def      \numu{\ensuremath{\nu_{\mu}}\ }

\def      \simge{\mathrel{%
   \rlap{\raise 0.511ex \hbox{$>$}}{\lower 0.511ex \hbox{$\sim$}}}}
\def      \simle{\mathrel{
   \rlap{\raise 0.511ex \hbox{$<$}}{\lower 0.511ex \hbox{$\sim$}}}}
\def      \BB{$\beta$-Beam\ }
\def      \SB{Super-Beam\ }

\title{Degeneracies at a $\beta$-Beam and a Super-Beam Facility}

\author{S. Rigolin\address[UAM]{Departamento de Fisica Teorica and 
                                Instituto de Fisica Teorica, \\ 
        Universidad Autonoma de Madrid, 28039 Madrid, Spain}%
        \thanks{The author acknowledges the financial support of MCYT through project 
        FPA2003-04597 and of the European Union through the networking activity BENE.}}       
\begin{document}

\begin{abstract}
The presence of degeneracies can considerably worsen the measure of the neutrino 
oscillation parameters $\theta_{13}$ and $\delta$. We study the physics reach of a 
specific ``CERN'' setup, using a standard $\beta$-Beam and Super-Beam facility. 
These facilities have a similar sensitivity in both parameters. Their combination 
does not provide any dramatic improvement as expected due to their almost identical 
L/E ratio. We analyse if adding the correspondent disappearance channels can help 
in reducing the effect of degeneracies in the $(\theta_{13},\delta)$ measure.
\vspace{1pc}
\end{abstract}
%
%
\maketitle
%
%
\section{Introduction}
\label{introduction}

The best way to simultaneously measure $(\theta_{13},\delta)$ is the (golden)
$\nue \!\!\raw \numu$ appearance channel \cite{Cervera:2000kp} (and its T and CP 
conjugate ones). Unfortunately this measure is severely affected by the presence 
of an eightfold degeneracy \cite{degeneracies}. 
Various methods have been considered to get rid of degeneracies: spectral 
analysis, combination of experiments and/or different channels. In principle, 
the eightfold degeneracy can be completely solved if a sufficient large set of 
independent informations is added. At the cost, of course, of increasing the 
number of detectors and/or beams and, consequently, budget needs. 

We try to understand here if the effect of degeneracies can be reduced adding 
informations from both the appearance and the disappearance channels at a single 
experiment. We consider as a reference two proposed CERN-based facilities: the 
standard-$\gamma$ \BB \cite{zucchelli} and the 4 MWatt SPL \SB \cite{jjsb}. 
Both neutrino beams are directed from CERN toward a 1 Mton water Cerenkov detector 
placed in the underground Fr\'ejus laboratory. The considered baseline is $L=130$ km. 
The average neutrino energy for both beams is $\approx 250$ MeV. 

The eightfold degeneracy for these two facilities has been comprehensively studied 
in \cite{Donini:2004hu,Donini:2004xx} and we refer to those papers for all the 
technical details and a complete set of bibliographic references. 
%
%
\section{$\beta$-Beam Appearance and Disappearance Channels}
%
%
The considered \BB setup consists of a $\bar \nu_e$-beam produced by the decay of 
$^6$He ions boosted at $\gamma = 60$ and of a $\nu_e$-beam produced in the decay of 
$^{18}$Ne ions boosted at $\gamma = 100$. The average neutrino energies of the 
$\nu_e,\bar \nu_e$ beams are 0.37 GeV and 0.23 GeV, respectively. 

\begin{figure}[t!]
\epsfig{file=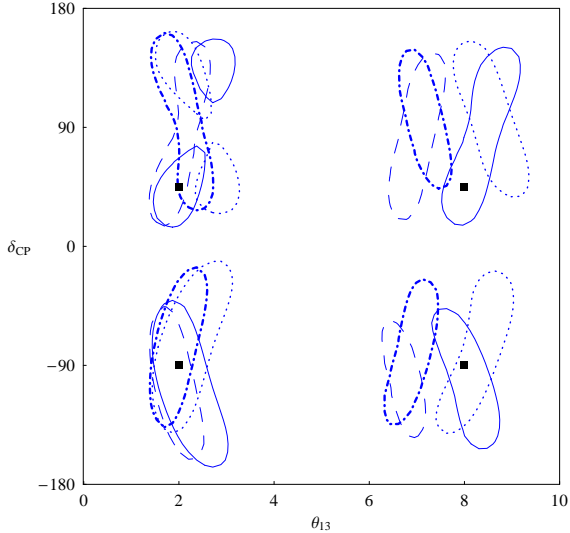,width=7.5cm,angle=0}
\caption{\it 
$90$\% CL contours in the ($\theta_{13},\delta$) plane using the appearance channel
after a $10$ years run at the considered \BB for $\theta_{13}= 2^\circ,8^\circ$, and 
$\delta= 45^\circ, -90^\circ$. A $5$\% systematic error is assumed. Continuous, dotted, 
dashed and dot-dashed lines stand respectively for the intrinsic, sign, octant and mixed 
degeneracy.}
\label{fig:appBB}
\end{figure}

The measurement of $(\theta_{13},\delta)$ at this facility has been already actively 
discussed in the literature \cite{allBB,Donini:2004hu}. In Fig.~\ref{fig:appBB} we plot 
our results for $\theta_{13} = 8^\circ$ and two different CP phases: $\delta = 45^\circ$ 
and $- 90^\circ$. The input value used in the fit is always shown as a filled 
black box. Throughout the paper we are using the following reference values for the 
atmospheric and solar parameters: $\Delta m^2_{atm}=2.5 \times 10^{-3}$ eV$^2$, 
$\theta_{23}=40^\circ$, $\theta_{12}=33^\circ$ and $\Delta m^2_{sol}=8.2 \times 10^{-5}$ 
eV$^2$. The $90$\% CL contours for each of the degenerate solutions are depicted in the 
plot assuming a $5$\% systematic error and are fully explained in the caption. 
%
%
In Fig.~\ref{fig:appBB} it can be seen the dramatic impact that degeneracies have 
in the precision of the measure of $(\theta_{13},\delta)$: 
(1) the error in the $\theta_{13}$ measurement is increased by a factor four (two) 
    for large (small) values of $\theta_{13}$\cite{Donini:2003vz}. The presence of 
    degeneracies has a small impact on the ultimate $\theta_{13}$ sensitivity; 
(2) the error in the $\delta$ measurement grows in a significant way in presence of the 
    clones, almost spanning half of the parameter space for small values of $\theta_{13}$. 
These facts are well understood: being the standard \BB a (short distance) counting 
experiment there are not enough independent informations to cancel any of the degeneracies. 

At the \BB the \nue disappearance channel is also available. The \nue disappearance 
probability does not depend on the CP violating phase $\delta$ and the atmospheric 
mixing angle $\theta_{23}$. Thus, the $\theta_{13}$ measurement, in this channel, is 
not affected by the intrinsic, octant and mixed ambiguities. 

%
\begin{figure}[t]
\epsfig{file=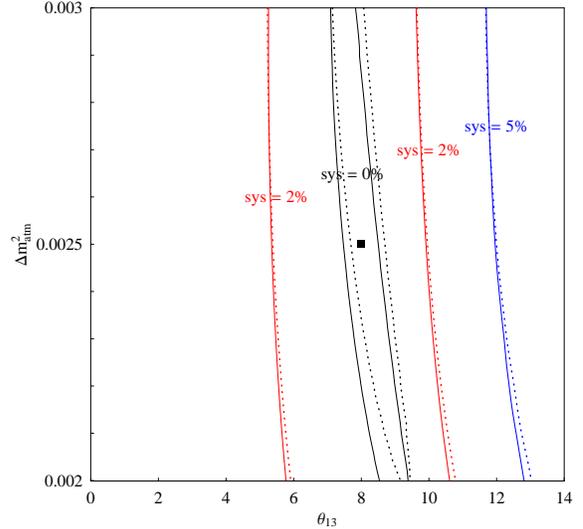,width=7.5cm,angle=0}
\caption{\it 
$90$\% CL contours in the ($\theta_{13},\Delta m^2_{atm}$) plane using the disappearance 
channel after a $10$ years run at the considered \BB for $\theta_{13}= 8^\circ$.
Three different values of the systematic errors have been considered: $0$\%, $2$\% and 
$5$\%. Continuous (dotted) lines stand for the true solution (sign degeneracy).}
\label{fig:disBB}
\end{figure}
%
In Fig.~\ref{fig:disBB} we present the \nue disappearance measure for three different 
systematic errors, namely $0$\% (``theoretical-unrealistic'' scenario), $2$\% 
(``optimistic'' scenario) and $5$\% (``pessimistic'' scenario). The $0$\% systematic 
line represent the ultimate (error free) reach of this experiment. The $2$\% and $5$\% 
lines will cover the optimistic and pessimistic feelings about future improvements in 
understanding a Mton water detector. The 90\% CL contours in the $(\theta_{13},
\Delta m^2_{atm})$ plane are shown for the input values $\theta_{13}=8^\circ$ and 
$\Delta m^2_{atm}=2.5 \times 10^{-3}$ eV$^2$. The \nue disappearance channel is only 
slightly sensitive to the sign clone. In fact, the dependence on the sign of the 
atmospheric mass difference arises only at ${\cal O}(\theta_{13}^2)$. As a consequence
the \nue disappearance channel is an almost ``clone-free'' environment for the \BB, 
as it is for reactors experiments. However, even in the case of an optimistic 2\% 
systematic error, no improvement is obtained adding the disappearance channel 
informations to the results of Fig.~\ref{fig:appBB} for the appearance channel. 
The resulting 90\% CL contours practically coincide with the previous ones, and for 
this reason we do not consider to present them in a separate figure. The $\theta_{13}$ 
indetermination coming from the clone presence in the appearance channel is smaller 
than the disappearance error itself. Only considering an unrealistic 0\% 
systematics the disappearance channel starts to be useful in eliminating clones.
%
%
\section{Super-Beam Appearance and Disappearance Channels}
%
%

%
\begin{figure}[t]
\epsfig{file=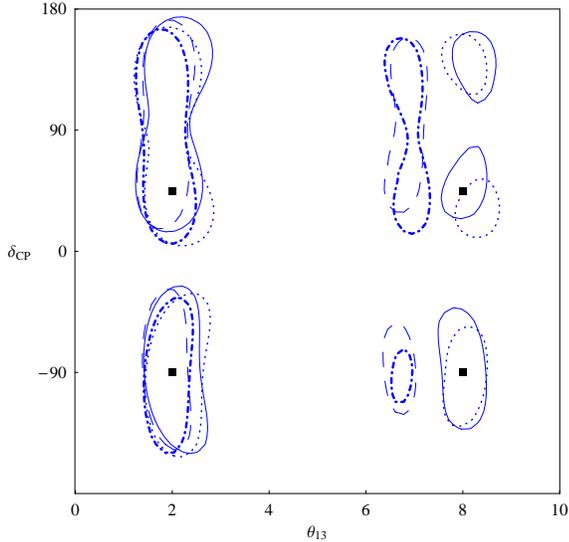,width=7.5cm,angle=0}
\caption{\it 
$90$\% CL contours using the appearance channel after a $2+8$ years run at the 
\SB for two different values of $\theta_{13} = 8^\circ$, and two values of $\delta$: 
$45^\circ$ and $-90^\circ$. A $5$\% systematic error is assumed. 
The legend is the same as in Fig.~\ref{fig:appBB}.} 
\label{fig:appSB}
\end{figure}
%
\begin{figure}[t!]
\epsfig{file=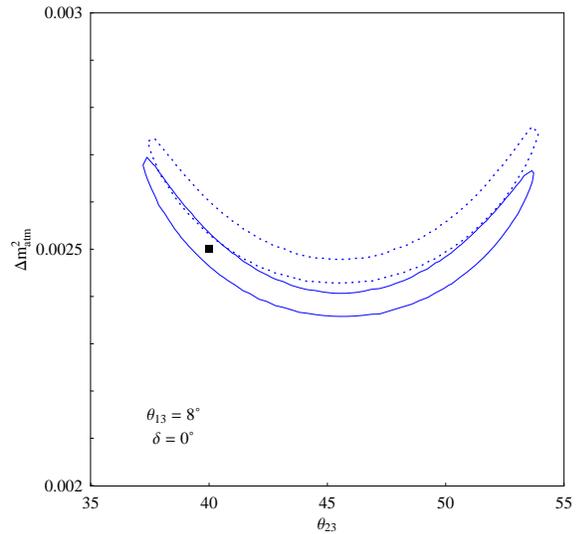,width=7.5cm,angle=0}
\caption{\it 
$90$\% CL contours in the $(\theta_{23},\Delta m^2_{atm})$ plane using the disappearance 
channel after a $2+8$ years run at the \SB for $\theta_{23} = 40^\circ$. A $2$\% systematic 
error is assumed. Continuous (dotted) lines stand for true (wrong) atmospheric mass sign 
assignment.}
\label{fig:disSB}
\end{figure}
%
The considered \SB setup is a conventional neutrino beam based on the 4 MWatt 
CERN SPL 2.2 GeV proton driver \cite{jjsb}. The average neutrino energies of the 
$\nu_\mu$, $\bar \nu_\mu$ beams are 0.27 GeV and 0.25 GeV, respectively. 
The possibility to measure $(\theta_{13},\delta)$ at a standard Super-Beam facility 
has been already widely discussed \cite{allSB,Donini:2004hu}. 

In Fig.~\ref{fig:appSB} we plot our results for $\theta_{13}=8^\circ$ and two different 
CP phases: $\delta = 45^\circ$ and $-90^\circ$. The input value used in the fit is 
shown as a filled black box. The $90$\% CL contours for each of the degenerate 
solutions are depicted in the plot assuming a $5$\% systematic error and are 
explained in the caption. 
As it appears from comparison of Fig. \ref{fig:appSB} with Fig. \ref{fig:appBB}, 
the ``figures of merit'' of a standard $\beta$-Beam and the SPL Super-Beam are 
very similar. Also the Super-Beam appearance channel is severely affected by 
proliferation of clones. The precision in measuring $(\theta_{13},\delta)$ is 
practically identical in the two cases. This is well explained by the comparable
statistics in the golden channel ($\nue \!\!\raw \numu\!\!$ vs $\numu \!\!\raw 
\nue\!\!$) and an almost equal $L/E$ ratio for the two experiments. There is no 
real synergy between this two setups and the only effect in summing these two 
experiments (concerning the $(\theta_{13},\delta)$ measure) is to double the 
statistics. 
%
\begin{figure}[t!]
\epsfig{file=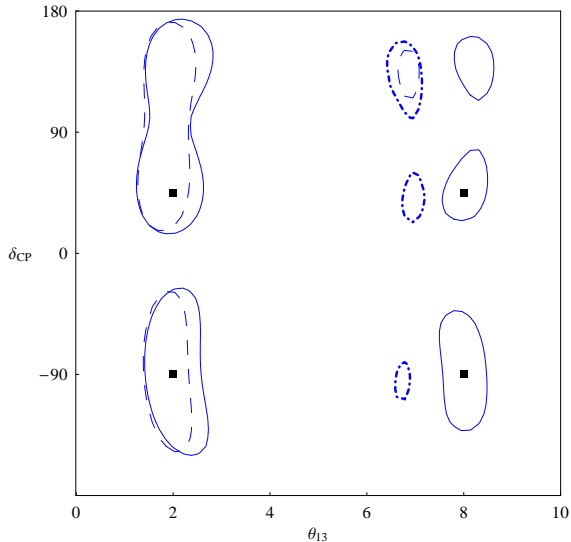,width=7.5cm,angle=0}
\caption{\it 
$90$\% CL contours using the appearance and the disappearance channels after a $2+8$ years 
run at the \SB, for $\theta_{13} = 2^\circ, 8^\circ$ and $\delta=45^\circ, -90^\circ$. 
A $5$\% ($2$\%) systematic error is assumed for the appearance (disappearance) channel. 
The legend is the same as in Fig.~\ref{fig:appBB}.} 
\label{fig:totSB}
\end{figure}
%

Nevertheless, the great advantage of the \SB facility compared with the \BB one 
is the possibility to measure directly the atmospheric parameters using the 
$\numu\!\!$ disappearance channel reducing, in particular, the atmospheric mass 
difference error to less than 10\%. In Fig. \ref{fig:disSB} we show the measure 
of $(\theta_{23},\Delta m^2_{atm})$ at the SPL \SB with a 2\% systematic error 
for a non-maximal atmospheric mixing, $\theta_{23} = 40^\circ$ and $\Delta m^2_{atm} 
= 2.5 \times 10^{-3}$ eV$^2$. The continuous (dotted) contour represents the fit 
to the right (wrong) choice of the sign of the atmospheric mass difference. Since 
we plot the results in the full $\theta_{23} \in [35^\circ-55^\circ]$ parameter 
space, the octant and mixed clones are automatically taken into account and do 
not appear as separate regions (within the considered errors). As one can notice 
the sign ambiguity implies that the errors on the atmospheric mass difference are 
roughly doubled with respect to what expected in the absence of degeneracies. 
The presence of degeneracies in the \numu disappearance channel can be easily 
understood looking at the the $\numu \!\! \raw \numu\!\!$ vacuum oscillation 
probability expanded to the second order $\theta_{13}$ and $(\Delta m^2_{sol} L/E)$ 
(see for example \cite{Akhmedov:2004ny,Donini:2004xx}).

In Fig.~\ref{fig:totSB} we present the simultaneous measurement of $(\theta_{13},\delta)$ 
using both the appearance (with a 5\% systematic error) and the disappearance (with a 
2\% systematic error) channels at the Super-Beam. Contrary to the $\beta$-Beam case, 
the disappearance \SB channel introduces in the fit significant changes. Notably enough, 
the sign clone has disappeared in any case considered. This is not a surprise as these 
fits are performed at a fixed $\Delta m^2_{atm}$: since in the disappearance channel 
the sign clone manifests itself at a larger value of $\Delta m^2_{atm}$ (see 
Fig.~\ref{fig:disSB}), in the combination with the appearance channel the tension 
between the two suffices to remove the unwanted clone in the $(\theta_{13},\delta)$ 
plane. Notice, moreover, that in some cases the octant clone is considerably reduced 
or even solved, due to the octant-asymmetric contributions in the \numu disappearance 
probability. Nonetheless, this does not mean that thanks to the combination of the 
appearance and the disappearance channels we are indeed able to measure the sign of 
the atmospheric mass difference. The mixed clones are generally still present for 
large values of $\theta_{13}$, thus preventing us from measuring the atmospheric mass 
difference sign if the $\theta_{23}$-octant is not known at the time the experiment 
takes place. It is clear that these results should be confirmed by a complete 
multi-dimensional analysis is underway. 
%
%
\section{Conclusions}
%
%
We have tried to understand the impact of disappearance measurements on the 
$(\theta_{13},\delta$) eightfold degeneracy for the standard-$\gamma$ \BB and the 
SPL Super-Beam. We presented a complete analysis of degenerations in the \nue and 
\numu disappearance channels: the \nue disappearance is affected only by a twofold 
degeneracy, since the \nue probability does not depend on $\delta$ and $\theta_{23}$. 
The \numu disappearance is affected by a fourfold-degeneracy. The inclusion of 
degeneracies almost double the error on the measure of $\Delta m^2_{atm}$.

The standard-$\gamma$ $\beta$-Beam setup looks somewhat limited, being the neutrino 
energy too low for using energy resolution techniques and the combination with the 
\nue disappearance, potentially of interest, is in practice useless once a realistic 
systematic error is taken into account. The \BB idea should be certainly pursued 
further, but using for example higher $\gamma$ options, where should be possible to 
take advantage of energy  resolution and, possibly, the silver channel. 

The SPL \SB appearance channel is also severely affected by degeneracies. However, in 
this case the complementarity between the appearance and disappearance channels can be 
fully exploited even when a realistic systematic error is taken into account. In 
particular, the sign ambiguity can be strongly reduced as the disappearance sign 
clone is located at a different $\Delta m^2_{atm}$. 
%
%

%
%
\end{document}